\documentclass[fleqn,twoside]{article}
\usepackage{amsmath}
\usepackage[headings]{espcrc2}

\readRCS
$Id: espcrc2.tex,v 1.2 2004/02/24 11:22:11 spepping Exp $
\usepackage{graphicx}
\usepackage[figuresright]{rotating}

\newcommand{\tdr}{\tilde{r}}

\newcommand{\prd}{\partial}

\newcommand{\beq}{\begin{equation}}
\newcommand{\eeq}{\end{equation}}
\newcommand{\bea}{\begin{eqnarray}}
\newcommand{\eea}{\end{eqnarray}}
\newcommand{\n}{\hspace*{-2.5mm}}

\newcommand{\ba}{\begin{array}}
\newcommand{\ea}{\end{array}}

\newcommand{\als}{\alpha_s}

\newcommand{\G}{\Gamma}
\newcommand{\g}{\gamma}

\newcommand{\bc}{\begin{center}}
\newcommand{\ec}{\end{center}}

\newcommand{\as}{a_s}

\def\bbuildrel#1_#2^#3%
{\mathrel{\mathop{\kern 0pt#1}\limits_{#2}^{#3}}}

\catcode`\@=11
\def\slash{\mathpalette\make@slash}
\def\make@slash#1#2{\setbox\z@\hbox{$#1#2$}%
  \hbox to 0pt{\hss$#1/$\hss\kern-\wd0}\box0}
\catcode`\@=12 

\def\nnb{\nonumber}

\newcommand{\ice}[1]{\relax}

\newcommand{\re}[1]{(\ref{#1})}

\title{
\vspace*{-3cm}
\centerline{\normalsize\hfill  SSFB/CPP-06-02}
\centerline{\normalsize\hfill  TTP06-03     }
\vspace*{1cm}
{}\thanks{
Presented by K.G.Ch.  at the  7-th International Symposium on Radiative Corrections, Shonan Village
(Japan), October 2005.}%
Multiloop Calculations: towards $R$
at Order  $ \alpha_s^4$
}

\author{P.~A.~Baikov\address{Institute of Nuclear Physics,
        Moscow State University, \\
        Moscow~~119992, Russia}%
       ,
        K.G. Chetyrkin\address[KUNI]{Institut f\"ur Theoretische Teilchenphysik, \\
        Universit\"at Karlsruhe, D-76128 Karlsruhe, Germany
        }%
\thanks{On leave from Institute for Nuclear Research
of the Russian Academy of Sciences, Moscow, 117312, Russia.}
        and
        J.H.~K\"uhn\addressmark[KUNI]}

\runtitle{}
\runauthor{}

\begin{document}

\begin{abstract}
We discuss recent developments  in multiloop calculations
aiming eventually in computing 
the total cross section for $e^+e^-$ annihilation into hadrons $\sigma_{had}$ 
in   order  ${{\cal O}(\alpha_s^4)}$.
\vspace{1pc}
\end{abstract}


\maketitle

\section{Introduction}
Grand Unification, the merging of the three gauge theories based on
$U(1),SU(2)$ and $SU(3)$ groups with their three independent coupling
$g_1$, $g_2$ and $g_3$ into the unified framework of  $SU(5)$ or, more
probable, $SO(10)$ gauge theory, is one of the most attractive
possibilities for physics beyond the Standard Model.  As is well
known, this can only be considered as attractive and natural option
consistent with the present values of the coupling constants  if
supersymmetry is realized in Nature.  Indeed, most of the models
predict that at least some supersymmetric partners of quarks,
leptons,  gauge and Higgs bosons will be discovered and studied in
detail at the next generation of colliders, LHS or ILC.

Once the masses of sufficiently many SUSY-partners are measured
precisely, the detailed energy dependence of the coupling constants
will be fixed and the test
for the unification can proceed with significantly improved
precision. In fact, it is conceivable that some hints about symmetry
breaking mechanism responsible for the transition from, say,  $SO(10$) to
$SU(1) \times SU(2) \times SU(3)$ could be inferred from such an analysis, in
particular, if combined with a similar study for Yukawa unification.

However, for such a program to be pursued successfully, the initial
values entering into the evolution equations---gauge couplings and masses
as measured at low energies---have to be known with  sufficient
precision. 
Of particular interest is the gain in resolution power from a more
precise determination of the strong coupling. A significant improvement 
is anticipated for the GIGA-Z option of the linear collider. 
Indeed, it has been argued in \cite{Aguilar-Saavedra:2001rg} that a sample of
$10^9$ Z decays might well lead to an experimental uncertainty of
$
\delta \alpha_s = 0.0008
{},
$
and even a value of $\delta \als = 0.0005$ has been  quoted \cite{Winter:2001av}.

The determination  of $\als$ from Z is either based on the  
measurement of the leptonic branching ratio
\[
R_{\ell}=  \Gamma_h/\Gamma_\ell=20.767\pm0.025
\]
which amounts to a simple counting of  leptonic and hadronic final
states,  or it exploits the leptonic peak cross-section 
\[
\sigma_\ell\sim\frac{\Gamma^2_\ell }{ \Gamma^2_{\mbox{tot}}}
=
    2.003\pm0.0027 \ \mbox{nb}
{}.
\]
In  the Standard Model fit  
\[
\alpha_s=0.1183\pm0.0030
\]
all this information is combined \cite{:2004qh}. In all these cases
the $\als$  dependence enters through the formula
\bea
\G_{\mbox{had}} = \Gamma_{0}\left(  
\rule{0mm}{4mm}
\right.
1\n
&+&\n \;\frac{\alpha_s}{\pi}
+1.409\;\frac{\alpha_s^2}{\pi^2}
\nonumber
\left.
-
12.767\;{\alpha_s^3\over\pi^3} 
\right)
\nonumber
\\
&+& \mbox{corrections}
\label{Zdecay:global}
\eea
where the second factor, conventually denoted as $1/3 R(s)$, gives the
perturbative series valid for massless quarks.  The ``corrections'' in the
above formula originate from quark mass terms $\sim  \frac{m_b^2}{M_Z^2}$,
from singlet terms which start in order $\alpha_s^2$, which  present in the axial
amplitude only and which originate from the imbalance between top and bottom quark
loops and, finally,  mixed, non-factorizeable QCD $\times$ electroweak corrections of
order $\als \, \alpha_{\rm weak}$.

All  these terms are discussed in \cite{ChKK:Report:1996}, and, indeed, the
``corrections'' terms in eq.~\re{Zdecay:global} are typically well under control.  The
dominant theory error one encounters in the extraction of $\als$
originates from estimates of the not yet available terms of order
$\als^4$. To estimate this uncertainty,  different strategies have been
advocated and we shall only list a small subset of the possible
choices. 

The most conservative estimate of the truncation error for an asymptotic 
series is identical to the last calculated term, presently of order 
$\als^3$    and leads to an uncertainty of  $\delta \als = 0.002 $
corresponding to $\delta \als/\als =  1.8 \%$.
Alternatively one may vary the renormalization scale which appears as 
argument of the strong  coupling  $\mu$ in a plausible range around $\sqrt{s}$ , say 
from $1/3 \sqrt{s}$ to $3 \sqrt{s}$. Using the correspondingly modified 
perturbative series to estimate $ \als (\mu)$ (and evolving $ \als(\mu)$ 
back to $ \als (M_Z)$) one obtains a quite asymmetric variation   
$\delta \als = {}^{+0.002}_{-0.0002}$ 
of the same order as estimated above. 

Various prescriptions  have been used to estimate not yet calculated higher order 
terms. From the Principe of Minimal Sensitivity (PMS)  \cite{Stevenson:1981vj} or 
Fastest Apparent  Convergence (FAC) \cite{Grunberg:1982fw} the forth order term is predicted as 
$-97 (\als/\pi)^4$ 
\cite{Kataev:1995vh}. This then would lead to a shift in $\als$ of $ 0.0006$, corresponding to 
$\delta \als/\als  = 0.5 \%$.

From this considerations it is evident that for a reliable determinations of  
$\als$ from GIGA-Z, fully exploiting the experimental precision, the evaluation of 
the $\als^4$ term  is a must. This arguments get even stronger in those 
cases, where low energy measurements are involved. 
The most extreme case in this direction is the determination of   $\als$ from 
the semileptonic versus leptonic decay rate of the tau lepton \cite{Davier:2005xq}.
Considering the enormous  events rates at low energy electron-positron storage rings, 
like $B$-  or charm-  or Tau- factories, R measurements at 10.5 GeV or at 3.7 GeV, just below the 
threshold for open bottom or charm production, could also lead to fairly 
precise measurements of  $\als$. Indeed, to achieve the benchmark 
precision 
Indeed, to achieve the benchmark 
precision 
$\delta \als(M_Z) = 0.003$
would require $\delta \als= .007$ at 10.5 GeV and 
$\delta \als = .013  $ at 3.7 GeV
\ice{

===========================================

E =10.5 

.1183 + .0030  = .1213

In[99]:=  AsRunDec[.1183,91.1876,10.5,4]

Out[99]= 0.176903

In[100]:=  AsRunDec[.1213,91.1876,10.5,4]

Out[100]= 0.183887

In[101]:= 0.183887 - 0.176903

Out[101]= 0.006984

E = 3.7

In[102]:=  AsRunDec[.1183,91.1876,3.7,4]

Out[102]= 0.235817

In[103]:=  AsRunDec[.1213,91.1876,3.7,4]

Out[103]= 0.248761

In[104]:=  0.248761 -  0.235817

Out[104]= 0.012944
}
Evidently only systematic 
uncertainties are the limiting factors for an ultra precise determination of
$\als$ . 
\section{The long march towards $R$ in $\als^4$}
Let us now briefly describe the strategy to evaluate the $\als^4$  term and 
the status of this calculation and a variety of physically and mathematically 
relevant results that are available at present. 

Consider the correlator  of two currents 
(all Lorentz and flavour indices are suppressed  
and quark masses are set to zero) 
$ j=\bar{\psi}\G\psi$ and $ j^{\dagger}=\bar{\psi}\G^{\dagger}\psi$ 
\[
\Pi^{jj}(q^2)  =
 i \int {\rm d} x e^{iqx}
\langle 0|T[
j(x) \, j^{\dagger}(0)]|0 \rangle
\]
which is related to the corresponding 
absorptive part  R(s) through
\[
R^{jj}(s) \approx  \mathrm{\Im} \,\Pi^{jj}(s- i\delta)
{}.
\]
Renormalized and bare  correlators are related through
\[
\Pi^{jj} = Z^{jj}  + Z^{j} \, \Pi^{jj}_B(q^2,\alpha_s^B)
\]
and the independence of $\Pi^B$  on the 
renormalization scale is reflected in the renormalization group equation  
of the form ($L = \ln \frac{\mu^2}{-q^2}$)
\beq
\left(
\frac{\prd}{\prd L}
- 2 \g^{j}(\als)
 +
\beta(\als)
\frac{\partial}{\partial \als}
\right)
           \Pi^{jj}
=  \g^{jj}(\als)
\label{RG}
{}.
\eeq	
Here  the anomalous dimension $\g^j(\als)$ is related to the corresponding 
renormalization constant as
\[
\g^j= \mu \frac{\prd}{\prd \mu} \ln (Z^j)|_{{}_{\scriptstyle \alpha_s^B}}
= \sum_{i >1} \g^j_i \left(\frac{\als}{\pi}\right)^i
{}
\]
while the QCD  beta-function
\[
\beta(\als) = \mu \frac{\prd}{\prd \mu} \als|_{{}_{\scriptstyle \alpha_s^B}}
=  - \als \sum_{i \ge 0 } \beta_i \left(\frac{\als}{\pi}\right)^{(i+1)}
{}.
\]
Equation \re{RG} clearly demonstrates that in order to find the $q^2$
dependent part of $\Pi$ at (N+1)-loop (corresponding to $\als^N$ order) one
needs the (N+1)-loop anomalous dimension $\g^{jj}$ and the N-loop
approximation of $\Pi$, (i.e. of order $\als^{N-1}$) including its constant
part.  Note that the operation $\als \frac{\partial}{\partial \als}$ raises the
power of $\als$ by one unit as $\beta(\als)$ starts from $\als^2$.
The problem of finding of a (N+1)-loop anomalous
dimension, however, can be reduced in a systematic automatized way to
the evaluation a proper combination of N-loop massless propagator integrals
\cite{gssql4,gvvq}.

The situation is significantly simplified whenever $\g^{jj}$ happens to
vanish.  One particular and physically relevant case is the $m_q^2/s$ term in
the small mass expansion of the absorptive part of the vector correlator,
where the corresponding four-loop ${\cal O}(\als^3)$ contribution to  $\Pi^{VV}$ 
is indeed sufficient to obtain the\ term of order ${\cal O}(\als^4 m_q^2/s)$
\cite{ChetKuhn90,Baikov:2004ku}.
\subsection{Reduction of massless propagators}
Thus,  to obtain the massless $R$-ratio in order $\als^4$, a large number of
four-loop massless propagators, including their finite parts, must be
evaluated. The complications can be best judged by contrasting the cases of
the three- and four-loop calculations: 11 (about 150) topologies with
and 3 (11)  topologies without insertions  are involved  in 
the three-(four-)loop case. 
The reduction has to be performed
to 6 (28)  master integrals. Most importantly, however, a set of recursion
relations based on integration by parts identities is available for the 3-loop
case.  These recursion relations have been constructed manually
\cite{ChT:ibp:1981} and implemented in the program MINCER \cite{Larin:1991fz}.

A straightforward implementation of this concept to the four-loop case seems
to be difficult at present. An alternative concept has been advocated in
\cite{Baikov:2003zq} which allows to perform the  reduction of an arbitrary
massless propagator to a sum of master integrals ``mechanically'' in the limit
of large space-time dimension $d$.  Consider the reduction of an amplitude
$f(d)$ which depends on the topology, the power of the various propagators and
the space-time dimension $d$ to master integrals:
\[
f(d)= \sum_{\alpha =\mbox{masters}} C^{\alpha}(d) \star M^{\alpha}(d)
{},
\]
where $M^{\alpha},$ with  $\alpha = 1-28$ stand for master integrals.  The
coefficients  $C^{(\alpha}(d)$ are known to be rational functions of
d; their determination is the central problem to be solved. In the
limit of large $d$  we have
\[C^{\alpha}(d)=\frac{P^n(d)}{Q^m(d)}  \bbuildrel{=\!=\!=}_{d \to \infty}^{}
\sum_k C^{\alpha}_k \ \ (1/d)^k
{}.
\]
The terms in the ${1/d}$ expansion can be expressed 
(with the use of the new representation for Feynman integrals developed in 
\cite{Baikov:tadpoles:96,Baikov:explit_solutions:97}) through simple 
Gaussian integrals---a task of purely algebraic nature.
Obviously, given sufficiently many coefficients, the
functions $C^{\alpha}$ can be reconstructed. However, the degrees m and n are
strongly dependent on the power of the propagators in the  integral 
under consideration with drastic effects on the effort involved in the evaluation.  
In the process of reduction a 4-loop diagram of
a typical topology  to Gaussian integrals
one  should  handle  a polynomial of 9 variables of degree $k$
consisting of
$
\frac{(9+k)!}{9! k!}
$
terms. For k=24 this corresponds to approximately $4 \cdot 10^7$ terms
or 4 GB storage space, while k=40 leads to $2 \cdot 10^9$ terms,
corresponding to 200 GB.
Weeks, partly even months of runtime and hundreds of GB disk space are 
required for the evaluation.
On the other hand, the purely algebraic   manipulations
required to perform the Gaussian integrals are ideally suited 
for treatment within the algebraic manipulation program FORM 
\cite{Vermaseren:2000nd} and  its parallel version PARFORM \cite{Fliegner:2000uy,Retey:2000nq}, 
which both are  specially tuned for dealing with a huge volume of algebraic
manipulations. The method of large $d$ expansion  has  been successfully applied
for solving a number of problems which will be briefly  discussed in the
next section.
\subsection{Important dots}
The reduction of the $(N+1)$ loop anomalous dimension to 
the calculation of $N$-loop massless propagators discussed above 
works ``natively''  only  for logarithmically  divergent integrals
while $\Pi^{jj}$ is  in general {\em quadratically} divergent. 
The only known way to apply the reduction in such cases is to
use (double!) differentiation w.r.t. the external
momentum $q$ to decrease the dimension.
This lead to "{dots}" $\equiv$
squared propagators which immensely complicate
all calculations.

An important and non-trivial simplification exists for the scalar (SS) correlator
due to the well-known Ward identity:
\begin{equation}
q_\mu q_\nu
\Pi^{{\rm V/A}}_{\mu\nu,ij}(q) =
(m_i \mp m_j)^2 \Pi^{\rm S/P}_{ij}(q)
+ {\cal O}(m_q^4)
{}.
\nonumber
\end{equation}
Basically this  means that the  ${\cal O}(m_q^2)$ part of the longitudinal part of
$VV$ correlator is identical to the massless $SS$ one. This allows to
compute the ${\cal O}(m_q^2)$ part of the $VV$ correlator instead of the
massless $SS$ which resulted in  diagrams with {\bf one squared propagator less}
and saves  a lot of work!
Unfortunately, no such simplification is known for the case of the
massless vector correlator.
\section{Results} 
Although the $R$ ratio to order $\als^4$ for the vector correlator is
not yet available a number of intermediate results of phenomenological
relevance have been obtained recently.


\noindent
$\bullet$
All 28 master integrals have been evaluated  analytically \cite{BCh:2006:masters}.

\noindent
$\bullet$
The terms of order $\alpha_s^4 n_f^3$  and  $\alpha_s^4 n_f^2$ 
for the $R$ ratio discussed in section 1   were obtained in \cite{Beneke:1992ch,ChBK:vv:as4nf2})
($a_s = \alpha_s/\pi$, dots stand for not yet computed term of order $n_f$ and
$n_f^0$)
\begin{align}
R(s) & = 3\,\Big\{
1 + a_s +
 a_s^2 \left( 1.986 - 0.1153 \ n_f
\right)
\nonumber
\\
&+ a_s^3
\left(
-6.6369 - 1.2001 \ n_f - 0.00518 \ n_f^2
\right)
\nnb
\\
\nonumber
&+ a_s^4
\left(
 0.02152 \ n_f^3 - 0.7974 \ n_f^2  +\dots
\right)
\Big\}
{}.
\label{RV:num}
\end{align}

\noindent
$\bullet$
The full dependence of the $\tau$-lepton    decay rate on the strange quark
mass $m_s$ up to order $\als^3$       has been obtained in \cite{Baikov:2004tk}.
\\
\noindent
$\bullet$
The quadratic term in the small quark mass expansion of the
$R$-ratio 
\[
R(s) = 3 \left\{r^V_0 +  \frac{m^2}{s} r^{V}_2 \right\}+\dots
{}.
\]
has been obtained in \cite{Baikov:2004ku}. For the number of active quark flavours $n_f=4$
the result reads
\bea
r^V_2/12 
 &=&  \as + 9.09722 \as^2
\nnb
\\
 &+& 52.913 \as^3
 + 128.499 \, \as^4
{},
\label{exact}
\eea
Let us mention that the $\as^4$   term is predicted to 177 and 193 by the estimates 
based on PMS and FAC respectively.
\\
\noindent
$\bullet$
Most recently the five loop anomalous dimension of the scalar correlator       
was obtained \cite{Baikov:2005rw}. In combination  with the finite part of the four-loop scalar 
correlator this corresponds to the R-ratio of the scalar correlator (denoted
with an extra tilde below)   and 
has important applications for QCD sum rules and for the Higgs decay rate to 
b-quarks \cite{Baikov:2005rw,Chetyrkin:2005kn}. 
Written for brevity in  numerical form   
only, the result reads:
\begin{align}
&\widetilde{R} =
1
+
5.6667  a_s
{+}
\left[
35.94
-1.359  \, n_f
\right]
a_s^2
\nonumber\\
&{+} a_s^3
\left[
164.1
-25.77  \, n_f
+0.259  \, n_f^2
\right] +
\label{RSnum}\\
&{} \,a_s^4
\left[
39.34
-220.9  \, n_f
+9.685  \, n_f^2
-0.02046  \, n_f^3
\right]
{}.
\nonumber
\end{align}
In Table \re{tab:FACPMS}  the  result for
$\tdr_4$ -- the coefficient in front of $\as^4$ term in \re{RSnum} -- 
is compared with predictions obtained in works
\cite{CheKniSir97,Chishtie:1998rz}.
Note  that the Principle of Minimal Sensitivity 
or the Principle of Fastest Apparent Convergence 
used in \cite{CheKniSir97} produce identical result at order $\alpha_s^4$.  The two
predictions of FAC/PMS for $\tdr_4$ correspond to either the consequence of
the prediction for the corresponding Euclidean quantity (second line) or to
the direct application of FAC/PMS to estimate $\tdr_4$ (the third line).  As a
consequence of the large cancellations in $\tdr_4$ the second prediction looks
much better than the first, despite the fact that the estimation of the
corresponding Euclidian coefficient  is quite close (within 10\%) to the exact
result (for more details see \cite{Baikov:2005rw}).
The  Asymptotic Pad\'e-Approximant Method (APAM) estimation of $\tdr_4$ 
constructed in \cite{Chishtie:1998rz}
fails to reproduce even the sign of
the exact result.
Finally,   predictions of the prescription proposed by Brodsky, Lepage and Mackenzie   (BLM)
\cite{Brodsky:1982gc} for the $n_f$ dependent terms of order $\alpha_s^4$ have been communicated
to the authors \footnote{M.~Binger and  S.~Brodsky, private communication.}:
$\as^4( -260 \, n_f + 13\, n_f^2  - 0.046\,  n_f^3)$
and are also in reasonable  agreement with the exact result of eq.~(\ref{RSnum}).
\begin{table}
\caption{\label{tab:FACPMS}Comparison
        of the results for  $\tdr_4$
        with earlier estimates based on
        PMS, FAC  and APAM.
         }
\begin{tabular}{cccc}
$n_f$                      & 3       & 4       &     5  \\
\hline\\
$\tdr_4$ (exact)                          & -536.8  &  -690.7   & -825.7  \\
$\tdr_4$ (\cite{CheKniSir97}, PMS, FAC)   & -945  &  -1099  & -1237 \\
$\tdr_4$ (\cite{CheKniSir97}, PMS, FAC)   & -528  &  -749  & -949 \\
$\tdr_4$ (\cite{Chishtie:1998rz}, APAM)   &  252    &  147      & 64  \\
\end{tabular}
\end{table}
\section{Conclusion}
The calculation of the $\widetilde{R}$-ratio of the scalar correlator at order
$\alpha_s^4$ has clearly demonstrated the enormous complexity inherent into
any calculation of such multiloop level in QCD.  The total CPU time
consumption amounts (very roughly) $3 \cdot 10^8$ seconds (about 10 years) if
normalized to the use of a stand-alone 1.5 GH PC.  Due to the heavy use of
the SGI cluster (of 32 parallel SMP CPU of 1.5 GH frequence each) the
calculation took about 15 calendar months.

The corresponding calculations for the vector correlator will increase this 
demand by another factor of  3 (optimistically) or 10 (pessimistically).
Clearly the combined experience of improved programs and better hardware 
will lead to the desired result within the  next few years.            

\ice{
\noindent
{\bf Acknowledgments}
\vspace{.4cm}
}
The work was supported by the Deutsche Forschungsgemeinschaft in the
Sonderforschungsbereich/Transregio SFB/TR-9 ``Computational Particle
Physics''.  The work of P. Baikov  was
supported in part by INTAS (grant 03-51-4007) and RFBR (grant 
05-02-17645).



\end{document}